\documentclass[twoside]{ahp}

\usepackage{zeh_h}

 \usepackage{graphicx}

\newcommand{\be}{\begin{equation}}
\newcommand{\ee}{\end{equation}}
\newcommand{\bea}{\begin{eqnarray}}
\newcommand{\eea}{\end{eqnarray}}

\begin{document}

\title{Roots and Fruits of Decoherence\protect\footnote{\ This version 2
of quant-ph/0512078 is slightly extended and revised compared to the
printed article in the proceedings of the Seminaire Poincar\'e of November
2005 (ed. by T. Damour, B. Duplantier and V. Rivasseau, Birkh\"auser
2006), while version 1 was identical with the proceedings.}}
 
\author{H. Dieter {\sc Zeh}\\
Universit\"at Heidelberg \\
www.zeh-hd.de}

\maketitle

\begin{abstract} The concept of decoherence is defined, and discussed in a
historical context. This is illustrated by some of its essential
consequences which may be relevant for the interpretation of quantum
theory.  Various aspects of the formalism are also reviewed for this
purpose.
\end{abstract}

\vskip.5cm

{\bf Contents}

1. Definition of concepts

2. Roots in nuclear physics

3. The quantum-to-classical transition

4. Quantum mechanics without observables

5. Rules versus tools

6. Nonlocality

7. Information loss (paradox?)

8. Dynamics of entanglement

9. Irreversibility

10. Concluding remarks

\section {Definition of concepts}

The concept of decoherence has become quite popular during the last two
decades. However, while its observable consequences have now been clearly
confirmed experimentally~\cite{Haroche,Zeilinger} (see also contributions
to this seminar), some misunderstandings
regarding its meaning seem to persist in the literature. The phenomenon
itself obviously does not depend on any particular interpretations of
quantum theory, although its relevance for them may vary
considerably~\cite{Schloss,Bac}. I am indeed surprised about the
indifference of most physicists regarding the potential consequences of
decoherence in this respect, since this concept arose as a by-product of
arguments favoring either a collapse of the wave function as part of its
dynamics, or an Everett-type interpretation. In contrast to the Copenhagen
interpretation, which insists on fundamental classical concepts, both
these interpretations regard the wave function as a complete and universal
representation of reality (cf.~\cite{ItOrBit}). 

So let me first emphasize that by decoherence I do neither just mean
the disappearance of spatial interference fringes in the statistical
distribution of measurement results, nor do I claim that decoherence
without additional assumptions is able to solve the infamous
measurement problem by {\it explaining} the stochastic nature of
measurements on the basis of a universal Schr\"odinger equation. Rather, I
mean no more (and no less) than the {\it dynamical dislocalization of
quantum mechanical superpositions}, which are defined in an abstract
Hilbert space with a local basis (given by particle positions and/or
spatial fields, for example). The ultimate nature of this
Hilbert space basis (the stage for a universal wave function) can only be
found in a unifying TOE (theory of everything), but does not have to be known
for the general arguments.

Dislocalization arises through the formation of entanglement of any system
under consideration (with states 
$\phi$) with another one, such as its unavoidable environment (described by
states
$\Phi$). This is often achieved by means of a von-Neumann type
``measurement'' interaction
\be
(\sum c_i \phi_i) \Phi_0 \to \sum c_i \phi_i \Phi_i \quad  ,
\ee
which would represent a logical {\it controlled-not} operation in the special
case
$i = 1,2$ and
$\Phi_0 = \Phi_1$. Ideal measurements, that is, those without recoil or
change of the state
$\phi_i$, define ``pure decoherence''. After the interaction, these
superpositions still {\it exist}, even though they {\it are not there} any
more~\cite{JZ,decoh}. The difference between these two, traditionally
equivalent phrases reflects the essential character of nonlocal
quantum reality. I am indeed convinced that the importance of decoherence
was overlooked for the first 60 years of quantum theory precisely because
entanglement was misunderstood as no more than a statistical correlation
between local objects (see Sect.~6).

Dislocalization of superpositions may be reversible  (``virtual'') or
irreversible in practice (``real'' decoherence). In the first case it would
either allow the complete relocalization of the superposition
(``recoherence''), or at least its reconstruction (the ``quantum erasure''
of measurement results). This distinction according to the
reversibility or irreversibility of decoherence {\it explains} also the
virtual versus real nature of other ``quantum events'', such as
radioactive decay, particle creation, or excitation. For example, decayed
systems remain in a superposition with their undecayed sources until
partial waves corresponding to different decay times are decohered from
one another. (This has the dynamical consequence of giving rise to an
{\it exact} exponential decay law -- see the contribution by Erich Joos
to these proceedings~\cite{Joos1}.) In contrast to recoherence (complete
reversal of the dislocalization, which would require a reversal of the arrow
of time), quantum erasure is compatible with the irreversible and non-unitary
dynamics of open systems -- related to a local entropy decrease at the cost
of an entropy increase in the environment~\cite{TD}.  

According to (1), dislocalization of superpositions describes a
distortion of the environment $\Phi$ by the system $\phi$ rather than a
distortion of the system by the environment (such as by classical
``noise''). This leads to the important consequence that
decoherence in quantum computers cannot be error-corrected for
in the usual manner by means of redundant information storage. Adding extra
physical quantum bits to achieve redundancy, as it would be appropriate to
correct spin or phase flips {\it in the system}, would in general even
raise the quantum computer's vulnerability against decoherence -- similarly as
the increased size of an object normally strengthens its classicality. (Error
correction codes proposed in the literature for this purpose are based on the
presumption of decoherence-free auxiliary qubits, which may not be very
realistic.)

In  special situations, decoherence may be observed as a
disappearance of spatial interference fringes. Only for mass points,
or center of mass positions of extended objects, are wave functions isomorphic
to {\it spatial} waves, and only after position measurements with
many equivalently prepared objects do they form a {\it statistical
distribution}. The interference pattern could then conceivable also have
been obscured by a slightly varying preparation procedure (for example
due to uncontrollable ``noise''), while decoherence according to Equ.~(1)
affects {\it individual} quantum states. Because of the latters'
nonlocality it  leads locally to a reduced density matrix that
describes formally an {\it apparent ensemble} of states (thus {\it not}
presuming it as in a statistical interpretation). The conceptually important
difference between true and apparent ensembles was clearly pointed out by
Bernard d'Espagnat~\cite{dEsp} when he distinguished between proper and
improper mixtures. In the case of virtual (reversible) decoherence, this
difference can even be observed as recoherence (a relocalization of the
superposition, that would be impossible for a proper mixture). 

Superpositions thus
define pure states, which characterize
{\it individual} properties that are not present in their formal
components. For example, the superposition of two different spinor states
is again an individual spinor state (up or down with respect to another
direction); the superposition of a $K$-meson and its antiparticle defines
a new particle ($K_{long}$ or $K_{short}$); that of a continuum of
positions (in the form of a plane wave) defines a certain ``momentum''
(wave number). Similarly, a superposition of products of the spin states of
two particles (even at different places) by means of Clebsch-Gordon
coefficients defines an individual state of total spin, while each
particle is then in an ``improper mixture'' because of its virtual
decoherence by the other one.  Under unitary transformations (described by
a Schr\"odinger equation) these total states remain pure and can never
become ensembles that might represent different measurement outcomes.
However, unitary decoherence may irreversibly lead to {\it apparent}
ensembles (improper mixtures) for local systems, which would precisely
explain the required ensembles of measurement outcomes if they were genuine
(proper). This consequence can hardly be an unrelated accident!  

\section {Roots in nuclear physics}

Nuclear physics provides some nice examples of many-particle systems
which are nonetheless clearly microscopic (found in energy eigenstates).
While I was involved in low energy nuclear physics during the sixties, I
became irritated by some methods which were quite successfully used. One of
them, called the time-dependent Hartree-Fock approximation, describes
``stationary'' states of heavy nuclei by means of
determinants of time-dependent single-nucleon wave functions. But how can
the mathematical solution of a static equation
$H\psi = E\psi$ know about a concept of time? Similarly,
certain deformed nuclei
were often described by means of a time-dependent ``cranking model'' in
order to calculate an effective moment of inertia, or to reproduce a
Coriolis type coupling between collective rotational states and individual
nucleons. However, both parameters characterize the spectra of static
energy eigenstates! It turned out that time is here used as a misleading
tool to describe {\it static superpositions} of one-parametric continua of
different determinants in order to construct quantum states for their
corresponding collective degrees of freedom (vibrations or rotations
around one axis, for example). 

For other collective modes, more than one
parameter may be required. General rotations, for example, have to be
represented by a non-Abelian symmetry group characterized by three Euler
angles. Superpositions then assume the form
\be
\Psi = \int d\Omega  f(\phi,\theta,\chi) U(\phi,\theta,\chi) \Phi
({\bf r}_1, \dots ,{\bf r}_n )
\quad ,
\ee
where $U(\phi,\theta,\chi)$ is the unitary transformation describing a
rotation and $d\Omega$ the volume element in this space, while $\Phi$ is a
deformed determinant or other ``model wave function''. There are many other
cases where entanglement is
classically circumscribed in terms of a time-dependent jargon. Well known
is the picture of ``vacuum fluctuations'' -- used to characterize a
static state of entangled quantum fields. 

 If a variational procedure 
\be 
\delta \langle \Phi | (H-E) |\Phi \rangle = 0 \quad ,
\ee 
with determinants $\Phi$
consisting of single nucleon wave functions
$\phi_i$,
 leads to a deformed solution
(as it happens for many heavy nuclei), one must first conclude
that
$\Phi$ can {\it not} be an approximation to the correct solution of $H\psi
= E \psi$, since it is far from being an angular momentum eigenstate.
However, using the degeneracy of these ``wrong'' solutions under rotations,
one may consider their superposition (2) as the next best step.
Simultaneous variation of the single-particle wave functions in
$\Phi$ and the superposition amplitudes
$f(\phi,\theta,\chi)$ then
 leads to angular momentum eigenstates and rotational spectra, including
Coriolis effects for the single particle motion~\cite{Z68}. 

The superposition (2) may be
regarded as being ``dislocalized'' over all nucleons in such a way that
they are all strongly entangled with one another.  A {\it strong} symmetry
violation of the model wave function $\Phi$ may be defined by the
quasi-orthogonality of slightly different orientations,
\be
 \langle \Phi | U(\phi,\theta,\chi) | \Phi
\rangle
\approx 0 \quad \rm{for} \quad {\it U} \neq 1 \quad ,
\ee
 as though the collective orientation
were an observable, and
$f(\phi,\theta,\chi)$ therefore the corresponding wave function. In a
similar way, phonon degrees of freedom {\it arise} in solid bodies. This
strong violation of rotational symmetry does not require a ``needle limit'' of
strong {\it geometric} asymmetry: it is a collective effect of many
slightly asymmetric single-particle wave functions (subsystems). For
product wave functions $\Phi = \prod_i \phi_i ({\bf r}_i)$, for example,
one would get
\bea
\langle \Phi |U(\phi,\theta,\chi)|\Phi \rangle =  \prod_i \langle \phi_i
|U(\phi,\theta,\chi)|\phi_i \rangle  =  \nonumber \\
\prod_i (1-\epsilon_i) 
\approx \prod_i \exp(-\epsilon_i) =
\exp(-\sum_i \epsilon_i ) \quad .
\eea
(For nucleons, their indistinguishability reduces this result somewhat, and
may let nuclei behave as a superfluid.) This quasi-orthogonality is very
similar to decoherence, which is often achieved by means of a product of
inner products of many stlightly disturbed environmental subsystems (such as
many scattered particles)~\cite{JZ}. In lowest approximation of the strong
symmetry violation, each nucleon in (2) ``feels'' only the deformed
(apparently oriented) self-consistent potential produced by the others. While
there is no {\it absolute} orientation in this case of rotational symmetry of
the exact Hamiltonian, the latter's dependence on inertial frames allows the
nucleons in higher order also to experience a Coriolis-type coupling with
the collective angular momentum.

So one may say that the individual nucleons ``observe'' an
apparent asymmetry in spite of the symmetric global superposition of all
orientations. A similar entangled superposition of different
pointer positions occurs in a quantum measurement that is described by von
Neumann's unitary interaction (1). This analogy led me to the
weird speculation about a nucleus that is big enough to contain a complex
subsystem which may resemble a registration device or even a conscious
observer. It/she/he would then become entangled with its/her/his
``relative world'', such as with a definite orientation. Does the
dynamical consequence described above then indicate a way to solve the
measurement problem? If the nucleons in the deformed nucleus dynamically
feel a definite orientation in spite of the global
superposition, would an internal observer then not similarly have to become
``aware of'' a {\it certain} measurement result?

This picture was also my first attempt towards a
(non-relativistic) quantum cosmology -- a kind of Everett interpretation
as I later discovered. When I learned about the static Wheeler-DeWitt
quantum universe, described by an equation
$H\psi = 0$, it also helped me to understand the concept of time as emerging
(cf.~\cite{Z8688} and Sect.~6.2.2 of Ref.~\cite{TD}). In contrast to a
macroscopic body, a nucleus in an energy eigenstate represents a closed
quantum ``universe''. However, it was absolutely impossible at that time to
discuss these ideas with colleagues, or even to publish them. An influential
Heidelberg Nobel prize winner frankly informed me that any further activities
on this subject would end my academic career!

Macroscopic objects are never found in energy eigenstates, but {\it always}
in states of certain (usually time-dependent) orientations or
positions. Therefore, one usually concluded that ``quantum theory
is not made for macroscopic objects'' or even the universe. According to
Niels Bohr, macroscopic systems have to be described in terms of
{\it presumed} classical (or ``every-day'') concepts -- even though they
would have to obey the uncertainty relations.

\section{The quantum-to-classical transition}

Much has been written about the quantum-to-classical
transition (cf.~\cite{ZurekToDay,decoh} and Zurek's contribution to these
proceedings). It is evidently crucial for a theory that describes reality
exclusively in terms of quantum states, while it would be of no more than
secondary importance (such as for explaining the absence of interference
patterns in scattering experiments) if classical concepts were {\it
presumed} for a probabilistic interpretation from the beginning.  I could
never accept such a fundamental divide between quantum and classical
concepts. So one has to understand the different {\it appearance} of
atoms, nuclei and small molecules on the one hand, and macroscopic objects
on the other. If both are described quantum mechanically, their energy
spectra differ quantitatively. For example, rotational states of
macroscopic objects are very dense. As a consequence, they cannot resist
entanglement with their environment even in the case of very weak
interactions. Their reduced density matrices must then always represent
``mixed states'', while  the locality of these interactions leads to the
vanishing of non-diagonal elements preferentially in the position or
``pointer'' representation. This is now called decoherence.

Although this term came up more
than ten years later (probably it was first used in talks given by
Gell-Mann and Hartle at the end of the eighties, preceding their
publication of 1993~\cite{GMH}), I had pointed out in a number of papers (see
\cite{diverse,KuZ}) that this disappearance of certain non-diagonal
elements of the density matrix explains superselection rules, which were
often postulated as fundamental restrictions of the superposition principle
(for example in axiomatic foundations of quantum theory). They were
assumed to hold for specific properties, such as electric charge, as well
as for ``classical observables'', although the axioms did not define a
precise boundary between quantum and classical concepts. 

In these early papers you will not even find
the word ``entanglement'' -- simply because this concept was so rarely
used at that time that I did not know this English translation of
Schr\"odinger's {\it Verschr\"ankung}. So I referred to it as ``quantum
correlations'' (in contrast to statistical correlations). Remember that even
Schr\"odinger, in his famous paper of 1935~\cite{Schr}, regarded {\it
Verschr\"ankung} as a mysterious probability relation (which would have
to characterize ensembles rather than individual states), since he  was
convinced that reality has to be defined in space and time. 

 However, what I had in mind went beyond what is now
called decoherence, since it was
inspired by the above mentioned picture of an observer inside a closed
quantum system. An external observer, who is {\it part of the environment}
of the observed object, becomes entangled, too, with the property he is
observing -- just as the observer within the deformed nucleus is
entangled with its orientation. He is thus part of
a much bigger ``nucleus'' (or closed system): the quantum universe. So he
``feels'', or can be aware, only of a definite value of the property he
has measured (or separately of different values in different ``Everett
worlds''). All you have to assume is that his {\it various} quantum states
which may exist as factor states in these different {\it components} of the
global wave function are the true carriers of awareness. This is even
plausible from a quite conventional point of view, since these decohered
component states, which are a consequence of the Schr\"odinger equation,
possess all properties required to define observers, such as complexity and
dynamical stability (memory). Indeed, these states are the same ones that
would arise in appropriate collapse theories if they were, according to von
Neumann's motivation, constructed in order to re-establish a
psycho-physical parallelism. But I do not see why such a modification, that
just eliminates all ``other'' components from reality, should be required. 

A genuine collapse that was simply triggered by irreversible
decoherence (as recently suggested by Roland
Omn\`es~\cite{Omnes}\footnote{The author has now informed me that he actually
meant to suggest a process of unitary recoherence. Because of the
irreversibility of decoherence, I do not expect such a hypothesis to be
realistic.}) would not even lead to {\it any} observable consequences, while
other models may either be experimentally confirmed or refuted. As long as
none of them is confirmed, it is just a matter of taste whether you apply
Occam's razor to the facts (by inventing new dynamical laws to cut off what
you cannot see) or to the laws (by leaving the Schr\"odinger equation
unchanged) -- although this choice must clearly have {\it cosmological}
consequences (such as a symmetric superposition of many different asymmetric
``worlds'' that might have been essential, that is, not yet decohered, during
early stages of the universe).

For me the most important fruit of decoherence (that is, of a universal
entanglement) is the fact that no classical concepts are required any more
on a fundamental level.  There is then also no need for a fundamental
concept of ``observables'', which would assume {\it certain} (classical)
values only upon measurement (see Chap.~4), and for uncertainty relations
restricting such values. The Fourier theorem for the wave function
explains this ``uncertainty'' in a natural way -- well known for classical
radio waves, which are themselves real and certain. When Bohr and
Heisenberg insisted that the uncertainty relations go beyond the Fourier
theorem, they were apparently thinking of {\it spatial} wave functions only
(thus neglecting entangled ones).

For microscopic objects which can be sufficiently isolated, the
experimental physicist has a choice between mutually exclusive
(``conjugate'') measurements, while macroscopic properties are decohered by
their unavoidable environment in a {\it general and specific} manner. This
explains their classical appearance. The corresponding quasi-classical
basis in Hilbert space then {\it appears} as a classical configuration
space, while the conventional ``quantization'' procedure may be regarded
as the re-introduction of these lost superpositions into the
(approximately valid) classical theory. Similarly, the classical world
appears local to us, since nonlocal entanglement becomes immediately
uncontrollable: it is decohered.

In order to illustrate the difference between this and the Copenhagen
interpretation, let me quote from a recent publication by Ulfbeck
and Aage Bohr from Copenhagen regarding the nature of {\it
quantum events}. They write~\cite{Ulfbeck}: ``No event takes place in the
source itself as a precursor of the click in the counter ...''. Hence,
there is no decay event in the atom, for example! So far I agree; this
conclusion, which is in contrast to earlier interpretations of quantum
theory, is enforced by experiments which use reflected decay fragments to
demonstrate recoherence (state vector revival) or interference with
partial waves resulting from later decay times.  In order to appreciate
this important change in the Copenhagen interpretation, one may compare
the {\it new} version with Pauli's claim from the fifties that ``the
appearance of a certain {\it position or momentum of a particle} is a
creation outside the laws of nature'' (my italics), which clearly refers
to the creation of particle properties (not just ``clicks'' in the counter).
Ulfbeck and Bohr then continue their sentence of above: ``... where the wave
function loses its meaning.'' Here I strongly disagree. After all, it is
precisely the arising uncontrollable entanglement with the environment,
described by a global wave function, which explains decoherence. These
authors are correct, though, when placing the creation of (apparent, I would
add) stochastic ``events'' in the apparatus, where the dislocalization of the
relevant superposition becomes irreversible FAPP (for all practical
purposes), thus creating an {\it apparent} ensemble of quasi-classical
wave packets. The dishonesty of the Copenhagen interpretation consists in
switching concepts on demand and regarding the (genuine or apparent)
collapse as a ``normal increase of information'' -- as though the wave
function represented no more than an ensemble of {\it possible} physical
states. However, this is ruled out by observed state vector revival
phenomena, for example.

Of course, you 
may {\it pragmatically} use classical concepts {\it as though} they were
fundamental -- even when studying decoherence as a phenomenon. One cannot
expect practicing physicists always to argue in terms of a universal
wave function. But they may keep in mind that there {\it is} a consistent
description (thus representing a ``quantum reality'') underlying their
classical terminology. Similarly, a high-energy physicist would use the
concepts of momentum and energy to describe
the objects in his laboratory, although he knows that  in
relativistic reality there is only a four-vector called
``momenergy''. Fortunately, there are other fruits of decoherence in the form
of observable phenomena which demonstrate decoherence in
action~\cite{Haroche,Zeilinger}. Nonetheless, the derivability in principle
of classical (such as particle) concepts undermines the motivation for the
Heisenberg picture as well as for Bohm's quantum mechanics. 

\section{Quantum mechanics without observables\protect\footnote{\ This
chapter is based on Sect.~2.2 of \cite{decoh}. 
It may be skipped for a quick reading.} }

In quantum theory, measurements are traditionally described by means of
``observables", which in the Heisenberg picture are assumed to replace
classical {\it variables}, and therefore have to carry the dynamical time
dependence. They are formally represented by hermitean operators,  and
introduced in addition to the concepts of quantum states and their
dynamics as a  fundamental and independent ingredient of quantum theory. 
However, even though often forming the starting point of a formal
quantization procedure, this ingredient may not be separately required if
physical states are universally described by general quantum states
(superpositions on an appropriate basis of states) and their dynamics. This
interpretation, to be further explained below, would comply with John
Bell's quest for a theory in terms of
``beables" rather than observables~\cite{Bell2}. It was for this reason
that his preference shifted from Bohm's theory to collapse models (where
wave functions are assumed to completely describe {\it reality}) during his
last years. (Another reason was his antipathy against the ``extravagance''
-- as he called it -- of the multiplicity of Everett worlds, which appears
in the form of myriads of empty components as well in Bohm's never
collapsing wave function.)

Let $|	\alpha\rangle$	be an arbitrary quantum state of a local system
(perhaps experimentally prepared by means of a ``filter'' or analyzer -- see
below). The {\it phenomenological} probability for finding this system in
another  quantum state
$|  n\rangle$, say, after an
appropriate measurement, is given by means of their inner product,
$p_n = |\langle n\mid \alpha\rangle |^2$, where both state vectors  are
assumed to be normalized. This state transition may either correspond
to a collapse or to a branching of the wave function -- although the states
of the apparatus and the environment are disregarded in this truncated
phenomenological formalism. The state
$|  n\rangle$ represents a specific measurement. In a position
measurement, for example, the number $n$ has to be replaced by the
continuous coordinates
$x,y,z$, leading to the ``improper" Hilbert  states $|\bf r \rangle$.
Measurements are called ``of the first kind'' or ``ideal'' if  the system
will  again be found in the state
$|  n\rangle$ (except for a phase factor) whenever the
measurement is  immediately repeated. {\it Preparations} of states can be
regarded as measurements which {\it select} a certain subset of outcomes
for further measurements.
$n$-preparations are therefore also called $n$-filters, since all
``not-$n$" results are thereby excluded from the subsequent experiment
proper. The above probabilities can be written in the form
$p_n = \langle \alpha\mid P_n\mid \alpha\rangle$, with  a special
``observable" $P_n := |  n\rangle \langle n| $, which is thus {\it
derived} from the kinematical concept of quantum {\it states} by using
their (phenomenological) probabilistic dynamics during measurements,
rather than being introduced as a fundamental concept.

Instead of these special ``$n$ or not-$n$ measurements" (for fixed $n$),
one can also  perform more general ``$n_1$ or $n_2$ or \dots\
measurements", with all
$n_i$'s mutually exclusive ($\langle n_i | n_j\rangle  =
\delta_{ij}$). If the states forming such a set $\{|  n\rangle \}$ are
pure and exhaustive (that is, complete, $\sum P_n = 1$),
they represent a basis of the corresponding Hilbert space. By introducing
an arbitrary ``measurement scale" $a_n$, one may construct {\it general}
observables
\be
A = \sum |  n\rangle a_n\langle n|  \quad , 
\ee
which permit the definition of
``expectation values"  
\be
\langle \alpha\mid A\mid \alpha\rangle  = \sum
p_na_n \quad .
\ee
  In the special case of a yes-no measurement, one has $a_n =
\delta_{nn_0}$, and expectation values become individual probabilities.
Finding the state
$ | n \rangle $ during a measurement is then also expressed as ``finding
the value $a_n$ of an observable".
A uniquely invertible change of scale,
$b_n = f(a_n)$, describes the  {\it same} physical  measurement; for
position measurements of a particle it would simply represent a
coordinate transformation. Even a measurement of the particle's potential
energy $V$
is equivalent to an (incomplete) position measurement if
the function
$V(\bf r )$ is {\it given}.

According to this definition, quantum expectation values must not be
understood as mean values in an ensemble that describes
ignorance about the precise state. Rather, they have to be interpreted  as
representing probabilities for {\it potentially arising} quantum
states
$|n\rangle$ -- regardless of the interpretation of this stochastic process.
If the set $\{|  n\rangle \}$ of such potential states forms a basis, any
state
$|  \alpha\rangle$  can be represented as a superposition
$|  \alpha\rangle  = \sum c_n|	n\rangle $. In general,  it neither forms
an $n_0$-state nor  any not-$n_0$ state. Its dependence on the complex
coefficients $c_n$ requires that formal states $|n\rangle$ which differ from
one another by a numerical factor must be different ``in reality''. This is
true even though they represent the same ``ray" in Hilbert space and
cannot, according to the measurement postulate, be distinguished
operationally. The states
$|  n_1\rangle	+ |  n_2\rangle $ and $|  n_1\rangle   - |  n_2\rangle $
could not be physically different from one another
   if $|	n_2\rangle $ and $-|  n_2\rangle $ were really the {\it same}
state. While operationally indistinguishable in the state
$\pm |n_2\rangle$ itself, any numerical factor -- such as a phase factor --
would become relevant in the case of recoherence. (Only a {\it global}
factor would thus be ``redundant''.) For this reason, projection operators
$|n\rangle\langle n|$ are insufficient to characterize
quantum states.

The expansion coefficients $c_n$ relating physically meaningful states --
for example those describing the various spin directions or different
versions of the K-meson -- have in principle to be determined (relative to
one another) by appropriate experiments.  However, they can often be
derived from a previously known (or conjectured) classical Hamiltonian
theory by means of ``quantization rules". In this case, the classical
configurations $q$ (such as particle positions or field variables) are
{\it postulated} to parametrize a basis in Hilbert space,
$\{|  q\rangle \}$,  while the canonical momenta $p$ parametrize another
one,
$\{|  p\rangle \}$. Their corresponding observables,
\be
Q = \int dq\,|  q\rangle q\langle q|  \quad  \rm{and} \quad
P = \int dp\,|  p\rangle p\langle p| \quad ,
\ee
are then required to obey commutation relations in analogy to
the classical Poisson brackets. In this way, they form an important
{\it tool} for constructing and interpreting the Hilbert space of
quantum states for this specific case. These commutators essentially determine
the unitary transformation
$\langle p\mid q\rangle $ (e.g. as a Fourier transform $\rm{e}^{\rm{i}
pq}$) -- thus more than what could be defined by means of the projection
operators
$|q\rangle \langle q |$ and  $|p\rangle \langle p |$. This algebraic
procedure is mathematically  very elegant and appealing, since the Poisson
brackets and commutators may represent generalized symmetry
transformations. However, the {\it concept} of observables (which form the
algebra) can be derived from the more fundamental one of state vectors and
their inner products, as described above.

Physical {\it states} are assumed to  vary in time in accordance with a
dynamical law -- in  quantum mechanics of the form
$\rm{i} \partial_t|  \alpha\rangle  = H|  \alpha\rangle $.  In contrast, a
measurement device is usually defined regardless of time. This must then
also hold for the observable representing it, or for its eigenbasis $\{
|  n\rangle \}$.  The probabilities
$p_n(t) = |\langle n\mid \alpha (t)\rangle |^2$ will therefore vary with
time according to the time-dependence of the  physical states $|
\alpha\rangle $. It is well known that this (Schr\"odinger) time
dependence is formally equivalent to the (inverse) time dependence of
observables (or the reference states
$|  n\rangle $). Since observables ``correspond" to classical  {\it
variables}, this time dependence appeared suggestive in the
Heisenberg--Born--Jordan algebraic approach to quantum theory.
However, the absence of {\it dynamical states} $|\alpha(t) \rangle$ from
this Heisenberg picture,  a consequence of insisting on {\it
classical} kinematical concepts, leads to paradoxes and
conceptual inconsistencies (complementarity, dualism, quantum logic,
quantum information, negative probabilities, and all that). The transition
to a Heisenberg picture essentially breaks down for open systems, which
are not dynamically described by a Hamiltonian.

An environment-induced superselection rule means that certain
superpositions are highly unstable against decoherence.  It is
then impossible in practice to construct measurement devices for
them. This {\it empirical} situation has led some physicists to
{\it deny the existence} of these superpositions and their corresponding
observables even in principle --  either by postulate or by formal
manipulations of dubious interpretation, often including infinities or
non-separable Hilbert spaces.

While {\it any} basis
$\{|  n\rangle \}$ in Hilbert space
defines formal probabilities,
$p_n =\break |\langle n | \alpha\rangle |^2$,  only a basis
consisting of states that are not immediately destroyed by decoherence
defines ``realizable observables". Since the latter usually form a genuine
subset of {\it all} formal observables (diagonalizable operators), they must
contain a nontrivial ``center" in algebraic terms. It consists of those
which commute with all the rest. Observables forming the center may be
regarded as ``classical", since they can be measured simultaneously with
{\it all} realizable ones. In the algebraic approach to quantum theory,
this center appears as part of its axiomatic structure~\cite{Jauch}.
However, since the condition of decoherence has to be considered
quantitatively (and may even vary to some extent with the specific nature
of the environment), this  algebraic classification remains an approximate
and dynamically emerging scheme.

These ``classical" observables thus characterize the
subspaces into which superpositions decohere. Hence, even if the
superposition of a right-handed and a left-handed chiral molecule, say,
{\it could} be prepared by means of an appropriate (very fast)
measurement of the first kind, it would be destroyed
before the measurement may be repeated for a test. In contrast, the chiral
states of all individual molecules in a bag of sugar are ``robust'' in a
normal environment, and thus retain this property {\it individually} over
time intervals which by far exceed thermal relaxation times. For discrete
states, this stability may even be increased by the quantum Zeno effect (see
\cite{joos} for a consistent and exhaustive discussion). Therefore, chirality
does not only appear classical in these cases, but also as an approximate
constant of the motion that has to be taken into account for defining
canonical ensembles for thermodynamics.

The above-used description of measurements of the first kind by means
of probabilities for transitions
$|  \alpha\rangle  \to |  n\rangle $  (or, for that matter,
by corresponding observables) is phenomenological. However, measurements
should be described {\it dynamically} as interactions between the measured
system and the measurement device. The observable (that is, the
measurement basis) should thus be derived from the corresponding
interaction Hamiltonian and the initial state of the  device.
As shown by von Neumann, this interaction must be
diagonal with respect to the  measurement basis
(see (1) and \cite{Zurek}). Its diagonal subsystem matrix  elements are
operators acting on the quantum state of the device in such a way
that the  ``pointer" evolves into a position appropriate for being read,
$|  n\rangle |	\Phi_0\rangle  \to |  n\rangle |  \Phi_n\rangle$. Here,
the first ket refers to the system, the second one to the device. The
states
$|  \Phi_n\rangle $, representing
different pointer positions, must approximately be mutually orthogonal,
and ``classical" in the explained sense.

Because of the dynamical version of the superposition
principle (that is, the conservation of superpositions in time), an initial
superposition
$\sum c_n|  n\rangle $ does {\it not} lead to definite pointer
positions (with their empirically observed  frequencies). If decoherence
is disregarded, one obtains their {\it entangled superposition}
$\sum c_n|  n\rangle |	\Phi_n\rangle $, that is, a state that is
different from all potential measurement outcomes $|n\rangle |\Phi_n
\rangle$. This dilemma represents the ``quantum measurement problem". Von
Neumann's interaction is nonetheless regarded as the first step of a
measurement (a ``pre-measurement''). Yet, a collapse still seems to be 
required -- now in the measurement device rather than in the microscopic
system. Because of the entanglement between system and apparatus, it would
then affect the total system.
    (Some authors seem to
have taken the phenomenological collapse in the {\it microscopic  system}
by itself too  literally, and therefore disregarded the state of the
measurement device in their measurement theory. 
Such an approach
is based on the assumption that quantum states always exist for all
systems. This would be in conflict with quantum nonlocality, even though it
may be in accordance with early interpretations of the quantum formalism.)

If, in a certain measurement, a whole subset of states $|n\rangle$ leads
to the same pointer position $| \Phi_{n_0} \rangle$, these
states can not be distinguished by this measurement. According to von
Neumann's interaction, the pointer state
$|  \Phi_{n_0}\rangle $ will now be
correlated with the {\it projection} of  the initial state onto the 
subspace spanned by this subset. A corresponding {\it collapse} was
therefore formally postulated  by L\"uders~\cite{Lueders} as a
generalization of von Neumann's ``first intervention" (as he had called the
collapse dynamics).

Since the interaction of a microscopic system with an appropriate measuring
device in this way {\it defines} the corresponding observable, the latter's
time dependence according to the Heisenberg picture
would then describe a formal time dependence of
the states {\it of the device}, though  paradoxically controlled by the
Hamiltonian {\it of the system}. This is a relic of the
formal ``correspondence'' between observables and classical {\it variables}.

The  question whether a certain formal observable (that is, a diagonalizable
operator) can be {\it physically realized} can only be  answered by
taking into account the unavoidable environment. A macroscopic
measurement device is {\it always} assumed to decohere into its
macroscopic pointer states. However, as mentioned in
Chapter 3, environment-induced decoherence by itself does not solve the
measurement problem, since the ``pointer states" $|
\Phi_n\rangle$ may be defined to include the total environment (the
``rest of the world").  Identifying the thus arising global
superposition with an {\it ensemble} of states (represented by a
statistical operator $\rho$) that leads to
the same  expectation values  $\langle A\rangle = \rm{tr} (A\rho)$
for a {\it limited} set of observables $\{A\} $ would
beg the question. This merely operational argument is nonetheless often
found in the literature.

\section{Rules versus tools}

As the Everett interpretation describes a ``branching quantum world'',
which mimics a collapsing wave function to the internal (co-branching)
observer, the question for the precise {\it rules} of this branching is often
raised -- similar to the dynamical rules for a collapse. Such collapse rules
would have to {\it postulate} the individual branches (including the ``pointer
states'') as well as their dynamical probabilities. In contrast,  decoherence
describes the branching by means of no more than a given Schr\"odinger
equation as a dislocalization of initially local superpositions. In this way,
the latter become gradually inaccessible to any local observer.
Decoherence neither defines nor explains
 the (conscious) observer as a specific subsystem. While the branching
is ultimately justified by the presumed locality of this observer, the
dislocalization itself is an {\it objective} dynamical process -- in
particular occcuring in measurement devices. 

This unitary dynamical process causes certain non-diagonal elements of the
reduced density matrices of all dynamically involved local systems (such as
those forming a chain of interactions which lead to an observation) to
approximately vanish. These {\it indicators} of dislocalized
superpositions are therefore often used to {\it define} decoherence. 
However, subsystems and their density matrices are no more than convenient
conceptual tools, useful because of the locality of all interactions and
the causal structure of our world (based on cosmic initial conditions that
are responsible for the arrow of time~\cite{TD}). In contradistinction to a
nonlocal superposition, the concept of a density matrix presumes the
probability interpretation. The degree of diagonalization of the reduced
density matrices may depend on the precise choice and boundaries of
subsystems, but this is irrelevant for a sufficient definition of
``macroscopically distinct'' global branches FAPP. Since irreversible
decoherence usually spreads without limit, the corresponding loss of phase
relations must be invariant under any finite enlargement of the decohered
{\it system}, while virtual, that is, local decoherence depends on the
choice of systems in a crucial way. Real decoherence may thus be regarded
as a {\it collapse without a collapse}.

While a genuine collapse theory would thus have to postulate (as part of the
dynamical law) probabilities for its various {\it possible} outcomes, in an
Everett interpretation {\it all} branches are assumed to remain in existence.
One can then meaningfully argue only about {\it frequencies} of outcomes
(such as the density of spots on a screen) in {\it series} of measurements
that have been performed in our branch. Graham was able to show more than
thirty years ago~\cite{Graham} that all those very abundant (by number)
``maverick Everett worlds'' which describe frequencies that are not in
accordance with the Born probabilities together possess a norm that vanishes
with increasing size of the series. While their exclusion is nonetheless {\it
not} a trivial assumption, the norm, which is usually chosen because of the
probability interpretation, plays here a similar r\^ole as phase space does
in classical statistical physics: it is dynamically conserved under the
Schr\"odinger equation, and thus an appropriate measure of probability.

\section{Nonlocality}

Let me continue with another reminiscence from the ``dark ages of
decoherence'' (that ended not before Wojcziech Zurek had published his
first papers on this subject in the Physical Review~\cite{Zurek}). After I
had completed the manuscript for my first paper on what is now called
decoherence~\cite{diverse}, the only well known physicist who responded to
it in a positive way for a long time was Eugene Wigner. He helped me to
get it published, and he also arranged for an invitation to a conference
on the foundations of quantum theory to be held at Varenna in 1970,
organized by Bernard d'Espagnat~\cite{Varenna}. 

When I arrived at Varenna, I found the participants (John Bell included) in
hot debates about the first experimental results regarding the Bell
inequalities, which had been published a few years before~\cite{Bell}. I
had never heard of them, but I could not quite share the general
excitement, since I was already entirely convinced that entanglement (and
hence nonlocality) was a well founded property of quantum states, which in
my opinion described reality rather than probability correlations. So I
expected that everybody would now soon agree with my conclusions. 

Obviously I was far too optimistic. Some physicists are
searching for loopholes in the experiments which confirm the violation of
these inequalities  until today -- even though all experimental results so far
were precisely predicted by quantum theory. Others (perhaps still the
majority) are interpreting nonlocality as a 
``spooky action at a distance'', which would have to affect tacitly
presumed local quantities or events (such as described by classical
concepts). I cannot see anything but prejudice (once shared by Einstein and
Schr\"odinger!) in such an assumption about reality. It is amazing that even
Bohm, who did assume the nonlocal wave function to be real, added classical
concepts to describe another (local) reality, which would then be able to
describe the evidently local observer, and for which the wave function acts as
no more than a pilot wave.
 
It appears strange, too, that certain ``measures of entanglement'' that
have recently been much in use~\cite{Peres} measure only reversible or
{\it usable} entanglement, while quite incorrectly regarding irreversible
entanglement (decoherence) as ``noise'' or ``distortion''. It is certainly
not an accident that this position appears related to Ulfbeck and Bohr's
above-mentioned statement. The observable consequences of Equ. (1)
demonstrate that quantum measurements can {\it not} be regarded as
describing a ``mere increase of information''  -- even in the absence of
any recoil.  Quantum measurement interactions produce {\it real} nonlocal
entanglement.

Locality here always refers to the common, three-dimensional space that
appears as the ``stage'' for classical physics. According to many modern (as
yet speculative) field theories it may be a subspace of some higher
dimensional space, which, just as three-space, occurs in turn as a space-like
foliation of some higher dimensional classical spacetime. Wave functions
are instead defined to live (and vary with time) in a very high dimensional
space that appears as a configurations space {\it of possibilities} only
after their superpositions have been eliminated by decoherence. This
``configuration space'' varies according to the degrees of freedom that are
relevant in a certain situation (defining a specific ``system'' -- such as
one apparently consisting of N particles). In quantum field theories, the
configuration space is more general -- consisting of all field amplitudes at
all points in space. Quantum theory is always {\it local on this stage} for
the wave function (which can be assumed to describe quantum reality). The wave
functions themselves form another space, {\it Hilbert space}, which thus
replaces configuration space or phase space to form the space of all
quantum states.

If reality is accepted to be {\it kinematically} nonlocal (in space), you
don't need any ``spooky teleportation'' in order to explain certain
experiments that appear particularly attractive to science fiction authors.
In all these experiments you have to {\it prepare in advance} a nonlocal
(entangled) state that contains, in one of its components,
precisely what is later claimed to be ported already at its final position.
For example, two spinors have to be prepared in the form
of a Bell state
\bea
&|\uparrow\rangle_A |\downarrow\rangle_B - |\downarrow \rangle_A
|\uparrow\rangle_B \nonumber = \\ &|\rightarrow \rangle_A |\leftarrow
\rangle_B - |
\leftarrow \rangle_A | \rightarrow \rangle_B = \dots \quad ,
\eea
where $A$ and $B$ refer to Alice's and Bob's place, respectively.
Nothing has to be ported any more when  Alice, say, follows the
usual ``teleportation'' protocol by measuring another
(local) Bell state that includes her spinor of (9) and the one to be
ported. 
Because of the ``real''
(irreversible) decoherence of the nonlocal superposition (9) caused by this
measurement, this {\it initial} Bell state becomes an apparent ensemble,
such that the entanglement it represents {\it appears} to be a statistical
correlation from the point of view of all local observers who do not yet know
the outcome (such as Bob). His apparently incomplete information may then be
``completed'' by apparently classical means (Alice sending a message to Bob).
In quantum terms, this ``information transfer'' means that Bob and his spinor
of (9), too, now become consistent members of the (partly irreversible) global
entanglement (see Joos's Sect.~3.4.2 of \cite{decoh}). This is then
experienced by Bob (in all his arising branches) as a collapse of the wave
function. If the relativistic universal Hamiltonian is local (an integral
over a spatial Hamiltonian density), it becomes obvious from this unitary
treatment that there can be no superluminal influence. Quantum nonlocality
is therefore compatible with the {\it dynamical locality} of quantum fields
that is often referred to as ``relativistic causality''. 

Alice assumes here the r\^ole of Wigner's well known
``friend'', who performs a measurement without immediately telling the
former the outcome (so when is the collapse?). If Pauli's remark of Chap.~5
were correct, though, something like telekinesis 
``outside the laws of nature'' would indeed have to create the
measurement result at Bob's place in the case of an initial nonlocal Bell
state. The term ``quantum information'' instead of entanglement is
therefore quite misleading: entanglement must be part of quantum reality
-- even though it may often become indistinguishable from a statistical
correlation in practice. 

Alice would need
a similar initial Bell-type superposition of the kind
\be
|CK \rangle_A |no CK\rangle_B - |no CK\rangle_A |CK\rangle_B 
\ee 
in order to ``beam'' Captain Kirk (CK) from her to Bob's place,
provided he {\it could} be shielded against decoherence until Alice either
``measures'' his absence at her place or, in the case of another outcome of
her measurment, sends Bob a message to perform a unitary transformation that
leads to the state
$|CK>$. (This hypothetical isolation of CK would permit the existence of a
local Schr\"odinger cat state
$\alpha |CK\rangle
\pm \beta |noCK\rangle$.) However, the Captain Kirk who is then found at Bob's
place could not be one who knows what happened at Alice's place {\it after}
the preparation of the initial Bell state. You would need a tremendously
more complex entangled state, that had to contain {\it all} later
``possibilities'' appropriately entangled in a nonlocal state, in order to
decide later {\it what} to ``beam''. The term ``quantum teleportation''
drastically misleads the public and should in my opinion not be used by
serious scientists. There is enough genuine weirdness, quantum theory has
to offer!

In practice, most nonlocal states (except for entanglement between very
weakly interacting subsystems, such as photons or neutral spins) would
immediately and irreversibly be further diclocalized in an uncontrollable
way. The apparent locality of our classical world is therefore the
consequence of its drastic nonlocality: classical ``facts'' (or events) {\it
appear} to be local, although they arise precisely by the dislocalization of
their superpositions.  Hypothetical local (classical) variables which have
{\it not} been measured are then often regarded as ``counterfactuals'' (but
not rejected as potentially appropriate concepts even though they {\it would}
contradict experimental results!), while the  nonlocal concepts
(superpositions) which successfully describe all these irreversible processes
dynamically (thereby {\it explaining} apparent facts) are simply dismissed as
``not conceivably representing reality''.

\section{Information loss (paradox?)}

The collapse of the wave function (without observing the outcome) or any
other {\it indeterministic} process would represent a dynamical
information loss, since a pure state is transformed into an ensemble of
possible states (described by a proper mixture, for example). The
dislocalization of quantum mechanical superpositions, on the other hand,
leads to an {\it apparent} information loss, since the relevant phase
relations merely become irrelevant for all practical purposes of local
observers. I will now argue that the ``information loss paradox of black
holes'' (Hawking's lost bet) is a consequence of decoherence, and {\it not} a
specific property of black holes. 

For a better understanding one may first consider irreversible processes in
classical mechanics, such as Boltzmann's molecular collisions (see
Chap.~9). Since they are based on deterministic dynamics (in analogy to
quantum unitarity), {\it ensemble entropy} must be conserved. However,
collisions lead to the formation of uncontrollable statistical
correlations, which are irrelevant FAPP in the
future. (They are important, though, for the correct backward dynamics
because of the specific cosmic initial condition that has to be assumed
for this Universe.) This apparent loss (namely, the dislocalization) of
{\it information} in this classical case affects {\it physical entropy},
since this entropy concept disregards {\it by definition} the arising
uncontrollable correlations~\cite{TD}. It is defined as an extensive
(additive) quantity, usually in terms of ``representative ensembles''
characterizing the local macroscopic variables, while microscopic (the
{\it real}) states -- including those of subsystems -- would be objectively
determined in principle by the global initial conditions because of the
presumed classical mechanical laws. In contrast, quantum mechanical
subsystems possess non-vanishing physical entropy (described by improper
mixtures) even for a completely defined global state. 

In general relativity (GR), ``information'' may
disappear when physical objects fall onto a spacetime singularity, but in
classical physics the real state of external matter would still exist and
remain determined. In contrast, for {\it quantum} mechanics on a classical
spacetime, the information loss would have to include all existing
entanglement with external matter, thus transforming the latter's improper
mixture into a proper one. This conclusion seems to remain true when the
black hole disappears by means of Hawking radiation, and this has been
regarded as a paradox, since it would violate global unitarity.

\begin{figure}[t]
\centering\includegraphics[width=.8\textwidth]{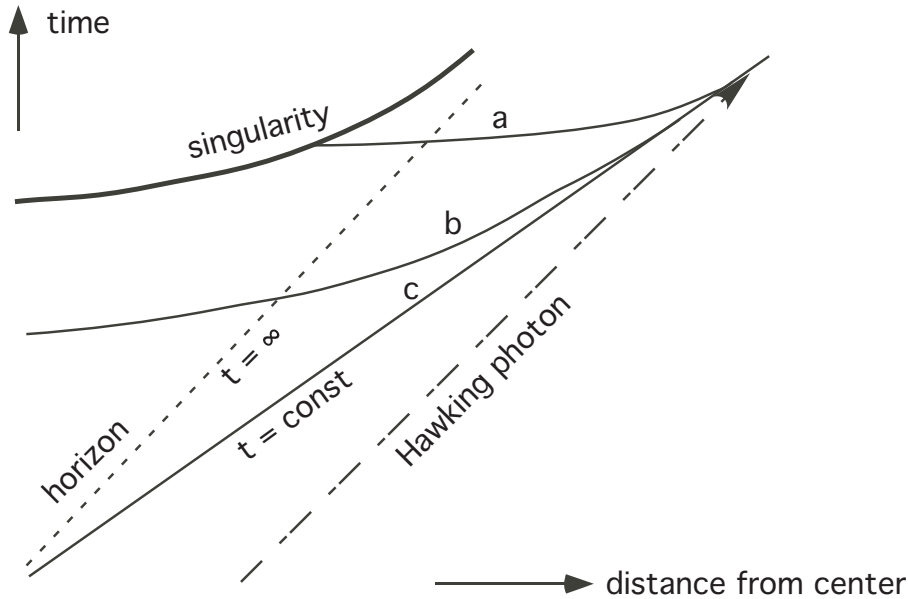}
\caption{{\footnotesize 
Various simultaneities for a spherical black hole in a
Kruskal type diagram: (a) hitting the singularity, (b) entering the regular
interior region only, (c) completely remaining outside (Schwarzschild time
coordinate $t$). Light cones open everywhere at $\pm45^0$
around the vertical time axis in this diagram, while lengths are strongly
distorted. Schwarzschild time is appropriate in particular for posing
external boundary conditions. The angle between the horizon and the line
$t= const$ can here be arbitrarily changed by a passive time translation.
This includes the (apparently close) vicinity of the horizon, which can
thus be arbitrarily ``blown up'' in the diagram -- thus transforming {\it
any} Schwarzschild time into the horizontal line $t = 0$, for example.}}
\end{figure}

One may consider the spacetime geometry of a black hole in
Kruskal-type coordinates (see Figure 1). Simultaneities used by external
observers in asymptotically flat spacetime (such as Minkowski time
coordinates in the black hole's rest system -- in the Figure appearing as
straight lines through the origin) can here be continued towards the center of
the spherical black hole in different, arbitrary ways. If everywhere chosen
according to the Schwarzschild time coordinate $t$, for example, they would
never intersect the horizon, but this choice does {\it not} affect the
density matrix representing the region far from the horizon (far right in the
Figure). The information loss noticed by an external observer can therefore
not be {\it caused} by the singularity -- no matter how long he waits. Not
even the horizon ever enters his past, and thus never becomes a ``fact'' for
him, while the Hawking radiation which he may observe would originate earlier
in Schwarzschild time than the horizon. Because of the diverging time dilation
in the close vicinity of the horizon, this region can can causally affect
only the very distant future, as superluminal effects are  excluded in spite
of quantum nonlocality (cf. Chap.~6).

On the other hand, a macroscopic black hole is permanently
affected by various kinds of decoherence~\cite{Kiefer} -- most importantly
by means of its retarded radiation. So this quantum radiation must be
highly entangled with the remaining black hole, and therefore with
radiation that is emitted later~\cite{Page}. If {\it usable} (macroscopic)
information about the black hole is stored in the external world (such as in
the emitted light), it defines separate Everett branches. While the
unitary dynamics determines the later global quantum state uniquely, it does
{\it not} determine an observer's branch: the present state of an observer
will, according to this unitary evolution, have {\it many} successors. Any
confirmation of the black hole's unitary dynamics would thus require the
complete recovery of  coherence, including the recombination of Everett
worlds -- just as it would be required to demonstrate unitarity for {\it all}
macroscopic objects. In practice, their evolution is irreversible ({\it not}
unitary). This means that the answer to Hawking's bet has nothing specifically
to do with black holes~\cite{BH}.

The spacetime metric with its event horizons and singularities may be
assumed to be ``real and certain'' only in classical GR. In quantum
gravity, even the spacetime geometry on which simultaneities are to be
defined has to be replaced by an entangled quantum state of matter and
geometry~\cite{J86,Z8688}. {\it All} macroscopic properties are thereby
decohered, and have to be associated with separate Everett worlds. The
Wheeler-DeWitt wave function $ \Psi[^3 G,
\phi_{matter}] $ (or its generalization to unified theories), which
describes their global superposition, defines ``probability amplitudes''
on {\it all} simultaneities (not just on those forming one geometro-dynamic
history, that is, a specific foliation of spacetime). 

This timeless wave
function has to obey certain {\it timeless} boundary conditions; it cannot
distinguish between past and future singularities -- such as
big bang and big crunch~\cite{Mazagon}. For example, these conditions may
exclude local singularities (or those with non-vanishing Weyl tensor), or
just any entanglement between them and regular regions. This would strongly
affect the wave function on all spatial geometries which intersect a black
hole horizon (perhaps even excluding the latter), and induce effects
corresponding to apparent final conditions. The WKB approximation, which
allows quasi-classical spacetime (hence proper times) to {\it emerge} by
means of the process of decoherence, may then completely break down in such
regions of configuration space~\cite{KZ}, while the classical spacetime
diagram of Figure~1 would lose its meaning close to the horizon.

However, this need {\it not} affect quasi-classical solutions restricted to
Schwarzschild simultaneities (the external part of the black hole). In
particular, Bekenstein's black hole entropy is a {\it general} result
derived in this region. It cannot be used to confirm {\it specific}
models of quantum gravity, such as string theory, although they may become
relevant in the internal region.

\section{Dynamics of entanglement}

The entangled state of any two quantum systems, if assumed to be pure, can
always be written as a single sum in the {\it Schmidt canonical
form}~\cite{Schmidt}
\be
\psi = \sum_i \sqrt{p_i}\phi_i \Phi_i \quad ,
\ee
where the states $\phi_i$ and $\Phi_i$ forming the two bases are {\it
determined} (up to linear combinations between degenerate coefficients) by
the total state
$\psi$. The coefficients can be chosen real and positive by an appropriate
choice of phases for the states forming the Schmidt bases, and have
therefore been written in the form
$\sqrt{p_i}$. In contrast to Equ. (1), the states $\Phi_i$ are now assumed
to be orthogonal: the expansion (11) is thus in general different from the
right hand side of (1). This Schmidt representation determines the
reduced density matrices in their diagonal form
\bea
\rho_\phi = \sum_i | \phi_i \rangle p_i \langle \phi_i | \quad , \nonumber
\\
\rho_\Phi = \sum_i | \Phi_i \rangle p_i \langle \Phi_i | \quad .
\eea
Since all systems must be assumed to be entangled with their
environments, the second system has in principle always to be
understood as the ``rest of the universe'' in order to represent a
realistic situation.

If the total state $\psi$ depends on time, the bases $\phi_i$,
$\Phi_i$ {\it and} the coefficients $\sqrt{p_i}$ must carry a separate
time dependence, which is determined, too, by that of the global state
$\psi (t)$. It can be explicitly written as~\cite{KuZ}  
\bea
{d \sqrt{p_i} \over dt} = Im \sum_{j} \sqrt{p_i} \langle ii|H | jj
\rangle  
\cr 
\rm i {d\phi_i \over dt} = \sum_{j \neq i}(p_i -
p_j)^{-1} \sum_m
\sqrt{p_m}[\sqrt{p_i} \langle ji | H | mm \rangle - \sqrt{p_i}\langle mm
| H | ij \rangle] \phi_i 
 \cr 
\rm{ i} {d\Phi_i \over dt}   =
\sum_{j \neq i}(p_i - p_j)^{-1} \sum_m
\sqrt{p_m}[\sqrt{p_i} \langle ij | H | mm \rangle - \sqrt{p_i}\langle mm
| H | ji \rangle ] \Phi_j   
\cr  + \sqrt{p_i} Re \sum_m \sqrt{p_m} \langle ii | H | mm \rangle \Phi_i
\quad .
\eea
The asymmetry between the two subsystems described by $\phi$ and $\Phi$ is
here due to an asymmetric phase convention,
and could be avoided by a different choice~\cite{Pearle}. 

In classical physics, subsystems would evolve deterministically,
controlled by time-dependent Hamiltonians which depend on the
state of the other system (thus forming coupled deterministic dynamics).
This classical concept of a time-dependent, environment-induced Hamiltonian is
often, quite inconsistently, also used in quantum mechanics -- for example in
the form of perturbing ``kicks'' or other kinds of ``noise'' instead of
genuine quantum interactions. However, an effective Hamiltonian induced by
the state of the environment would require a separately existing state for
the latter. In contrast,  Equs. (13) define highly nontrivial (hardly
practically usable) {\it nonunitary} subsystem dynamics. For this reason, the
``probabilities''
$p_i$ and the entropy
$\sum p_i \ln p_i$ defined by them must generically change with time. In
particular, initially separating systems will usually become entangled.

Although Equs. (13) define a continuous evolution for each
term of the Schmidt representation, this dynamics seems to become singular
whenever two diagonal elements of the density matrix, $p_i$ and $p_j$,, say,
become equal. Closer inspection of the dynamics reveals that two eigenvalues
coming close repel each other (unless the corresponding matrix elements of
the Hamiltonian vanish exactly), and therefore never intersect as
functions of time (see Figure~2). Thereby, the quasi-singular dynamics
(13) of the states forces the latter to interchange their identity within
a very short time. In other words: degenerate probabilities (such as required
in Bell states!) can never occur exactly, while the formal continuity of
Schmidt components in Figure 2 is entirely unphysical (not representing any
preserved identity of states). Subsystem density matrices are {\it not}
affected by this phenomenon, since the resonance terms are a consequence of
the ambiguity of their degenerate eigenstates. The non-unitary dynamics of
entangled density matrices can implicitly (that is, depending on the
solutions of (13)) be written as~\cite{Z73} 
\bea  
i  {d\rho_\Phi \over dt} &&:= i  {d\sum p_i \Phi_i \Phi_i^\ast \over
dt}
\nonumber \\ &&=
\sum_{i,j}\left( \sqrt{p_i}\langle ij|H| \psi \rangle
-\sqrt{p_j}\langle
\psi |H| ji \rangle
\right) \Phi_i \Phi_j^\ast \quad .
\eea

\begin{figure}[t]
\centering\includegraphics[width=.7\textwidth]{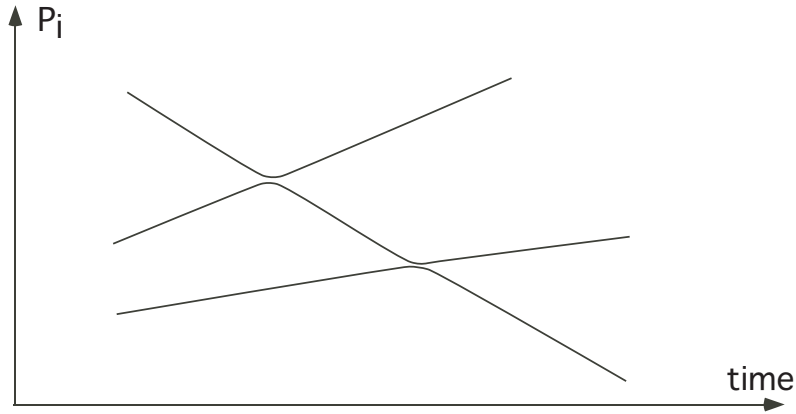}
\caption{{\footnotesize Trajectories of different probabilities $p_i(t)$
repel each other, while their corresponding factorizing Schmidt components
interchange their identity (including all memories). Causal histories of
Schmidt components thus intersect in this diagram, although they
never touch. This effect may also be regarded as a pure artifact of the
Schmidt representation.}}
\end{figure}

Initially separating
(factorizing) states are of special interest for the process of
decoherence. While this assumption enforces an initial degeneracy, though only
for between vanishing probabilities, the initial component with
$p_0 = 1$ can not linearly depend on time because of the time
reversal symmetry of the global Schr\"odinger equation. In this small-times
approximation, its precise form can be derived by means of perturbation
theory as
\be
p_0(t) \approx 1 - t^2 A \quad ,
\ee
where the quantity
\be
A = \sum_{j\neq 0, m \neq 0} |\langle jm | H | 00 \rangle | ^2 
\ee
has been called a {\it deseparation parameter}~\cite{KZ}. It measures the
rate of arising entanglement (that is, the growing deviation from a separating
state).  Index pairs $jm$ here refer to product states $\phi_j \Phi_m$. Note
that
$A$ is different from the quantity 
\be
B = \sum_{jm \neq 00}  |\langle jm | H | 00 \rangle | ^2 \ge A \quad ,
\ee
which measures the {\it total} change of the global state in this
approximation (including the ``classical'' change characterizing  a
time-dependent product). If the environment and the interaction Hamiltonian
$H$ are given, the deseparation parameter $A$ can be used to estimate the
robustness of certain states against decoherence. For example, coupled
harmonic oscillators turn out to be robust when in coherent states (such
as in states describing classical fields), while their energy eigenstates
(such as photon number eigenstates) are unstable against
decoherence~\cite{KuZ} (a result that was derived again in
\cite{ZkCoh}). 

\section{Irreversibility}

The dynamics of entanglement, derived from a global Schr\"odinger
equation and discussed in the preceding chapter, is time reversal
invariant. An asymmetry may be introduced by assuming {\it
initially}
 separating states, for example when using Equ. (15)
exclusively for
$t > 0$. However, if one of the two systems is indeed the ``rest of the
universe'', this assumption can be exact only once (such as for all local
systems at the big bang), although similar conditions may then {\it
approximately} occur later, too, at least within dynamically autonomous
Everett components after they have branched off (or, alternatively, after a
time-asymmetric collapse). 

A similar situation of special initial conditions is known to be required
for irreversible processes in general. For example, Boltzmann
collisions are assumed to affect initially uncorrelated single particle
distributions (defined in
$\mu$-space), although correlations must have built up ever since the big
bang. The point here is that these correlations remain irrelevant in the
future because of the chaotic nature of these classical systems and the
enormous lengths of their Poincar\' e recurrence times -- reflecting a very
large information capacity of correlations. Decoherence produced by
scattering of photons or molecules~\cite{JZ} is analogous to
Boltzmann's entropy production by means of molecular collisions, while
coupled oscillators, which have often been used to study
decoherence~\cite{CD}, may be useful because of their analytical
solutions, but are known to
 possess pathological properties: they are not {\it mixing} in a
thermodynamical sense if treated as closed systems. Similar arguments apply
to closed spin lattices, which may therefore better be
regarded as representing {\it virtual} decoherence -- in strong contrast
to systems which would classically show chaotic behavior~\cite{ZP}.

In quantum theory, reversibility would  not even hold in principle if a
genuine stochastic collapse of the wave function were assumed.  If the
Schr\"odinger equation is instead assumed to be universally exact,
recoherence of different Everett branches would have to be taken into account
in order to facilitate reversibility. This can be excluded FAPP (just as the
reversal of any other macroscopic arrow of time), and it is therefore
neglected in the usual description of phenomena by means of irreversible
master equations, for example.  

This general mechanism of irreversible dynamics was elegantly formulated by
Zwanzig~\cite{Zwanzig} by means of idempotent operators $P$
acting either on classical statistical probability distributions or on
density matrices
$\rho$. These ``Zwanzig projectors'' need neither be linear nor hermitean.
They are chosen to formalize the neglect of some kind of
``irrelevant information'' $\rho_{irrel} = (1-P)\rho$. Idempotence ($P^2 = P$)
means that throwing away the same information twice at the same time does not
add anything to doing it once. One should keep in mind, however, that density
matrices usually represent improper mixtures, such that a {\it dynamically
applied} Zwanzig projection would include collapse dynamics. 

 A linear Zwanzig projection is given by
tracing out the environment,
\be
\rho_{rel} = P_{sub}\rho = \rho^{(\phi)} = \rm{Trace}_\Phi \rho \quad ,
\ee
while a nonlinear one may merely neglect correlations between the two
subsystems,
\be
\rho_{rel} = P_{sep}\rho = \rho^{(\phi)} \rho^{(\Phi)} = \rm{Trace}_\Phi
\rho \,
\rm{Trace}_\phi \rho \quad .
\ee
There are many other applications of this very general concept,
which can also be regarded as a {\it generalized coarse graining}. For
example, one may neglect quantum correlations (entanglement), while
leaving all classical correlations intact. As another example, Boltzmann's
{\it Sto\ss zahlansatz} neglects dynamically all arising statistical
correlations between particles. 

\begin{figure}[t]
\centering\includegraphics[width=.8\textwidth]{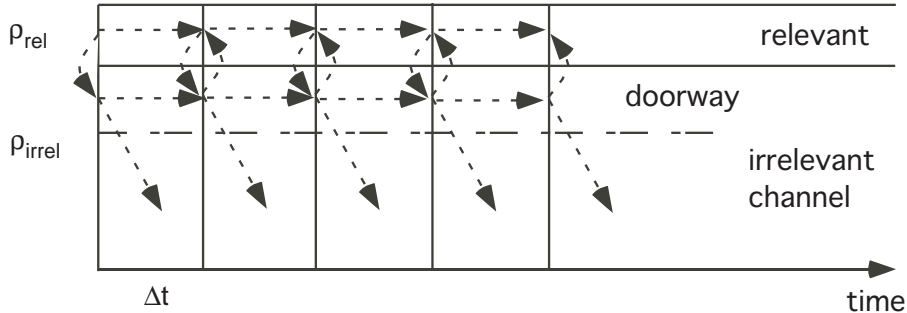}
\caption{{\footnotesize Propagation of ``information'' contained
in the statistical distribution or density matrix $\rho$ for the case
$\rho_{irrel}(t=0) = 0$. It propagates through two coupled ``channels''
(relevant and irrelevant -- see also Figs.~3.2 - 3.4 of \cite{TD}).
Intervals $\Delta t$ characterize steps of integration for the coupling
$PH(1-P) + (1-P)HP$ in the interaction representation.
The entropy, defined in terms of
$\rho_{rel}$, grows irreversibly FAPP if relevant information
disappears into the irrelvant channel for very long Poncar\' e times
(usually by far exceeding the age of the universe). This may allow the
formulation of master equations for
$\rho_{rel}$. If the irrelvant channel describes entanglement, the
irreversible loss represents real decoherence, while the ``doorway
channel'', which is part of the irrelvant channel, corresponds to virtual
decoherence. }}
\end{figure}

Zwanzig projectors are useful, in particular, since they often allow the
formulation of an effective (approximately autonomous) dynamics for
$\rho_{rel}$. This requires the special initial, but approximately
maintained, condition
$\rho_{irrel} \approx 0 $, which may lead to a master equation of the form
\be
{d\rho_{rel} \over dt} = - G_{ret} \rho_{rel} \quad ,
\ee
applicable in the forward direction of time. $G_{rel}$ must be a positive
operator, acting on the density matrix, in order to describe an entropy
increase (loss of information). The entropy is here defined in terms of
$\rho_{rel}$ as 
\be
S = -k_B \rm{Trace}[\rho_{rel} \ln
\rho_{rel}] \quad .
\ee 
This dynamics is schematically depicted in Figure 3  (cf.~Chap.~3 of
\cite{TD}), where relevant information is permanently lost to the
``irrelevant channel''. For the Zwanzig projector $P_{sep}$ this process
describes the production of entropy by means of decoherence, while for
$P_{sub}$ it would include an information transfer from the system to the
environment (that is, an entropy transfer from the environment to the
system -- usually attributed to heat flow or noise).  The ``doorway channel''
in this picture describes irrelevant degrees of freedom which can directly
interact with the relevant channel. In the case of entanglement, it
corresponds to virtual decoherence. Note, however,  that, for these quantum
Zwanzig projectors, $\rho_{rel}$
includes an apparent (improper) ensemble of different Everett ``worlds'',
since their superposition has been dislocalized (become irrelevant to
local observers). This means that the formal entropy
$S[\rho_{rel}]$ contains not only physical entropy, but also the ``entropy
of lacking information'' about the outcomes of all past measurements or
spontaneous symmetry breakings. Physical entropy is defined as a {\it
function} of these macroscopic quantities. Definite macroscopic
(classical) histories in quantum mechanical description are thus based on
a stochastic collapse or branching of the wave function, while classically
 they merely require the {\it selection of subensembles},
which represent incomplete information (``ignorance'') about the real
physical states. Such a selection is required in order to describe
individual macroscopic histories whenever microscopic causes have
macroscopic effects. In quantum theory, this ``selection'' corresponds to a
genuine quantum measurement (collapse or branching).

Explicit models for the irreversible process of decoherence and their
consequences are discussed in the accompanying paper by Erich
Joos~\cite{Joos1}.

\section{Concluding remarks}

To conclude, let me emphasize that the concept of decoherence does not
contain any new physical laws or assumptions beyond the established
framework of quantum theory. Rather, it is is a consequence of the
universal application of quantum concepts (superpositions) and their
unitary dynamics.

However, a consistent interpretation  of this theory in accordance with the
observed world requires a {\it novel and nontrivial identification of
observers} with appropriate quantum states of local systems which exist only
in certain, dynamically autonomous {\it components} of the global wave
function. Accordingly, it is the observer who ``splits''
indeterministically -- not the (quantum) world.

This interpretation is an attempt to replace the ``pragmatic
irrationalism'' that is common
in quantum theory textbooks (complementarity, dualism, fundamental uncertainty
etc.) by a consistent application of just those concepts which are actually,
and without exception {\it successfully}, used when the theory
is applied.

\vskip.8cm

\noindent {\bf Acknowledgements:} I wish to thank Erich Joos and Claus
Kiefer for their collaboration over many years,  for their persisting
interest in the subject, and for their critical comments on the
manuscript of this contribution.

\end{document}